\documentclass[12pt]{iopart}
\usepackage{iopams,amssymb}
\newcommand{\dket}[1]{| \, #1 \rangle\!\rangle}
\newcommand{\dbra}[1]{\langle\!\langle #1 \, |}  
\begin{document}

\title[Informationally complete measurements and groups
representation]{Informationally complete measurements and groups
representation}

\author{G. M. D'Ariano, P. Perinotti, and M. F. Sacchi\footnote[3]{To
whom correspondence should be addressed (msacchi@unipv.it)}}

\address{Unit\`a INFM and Dipartimento di Fisica ``A. Volta'', \\
 Universit\`a  di Pavia, via A. Bassi 6, I-27100 Pavia, Italy}

\begin{abstract}
Informationally complete measurements on a quantum system allow to
estimate the expectation value of any arbitrary operator by just
averaging functions of the experimental outcomes.  We show that such
kind of measurements can be achieved through positive-operator valued
measures (POVM's) related to unitary irreducible representations of a
group on the Hilbert space of the system.  With the help of frame
theory we provide a constructive way to evaluate the data-processing
function for arbitrary operators.
\end{abstract}


\section{Introduction}
The aim of any measurement is to retrieve information on the state of
a physical system. In classical mechanics, measuring the location on
the phase space provides a complete information on the system. On the
other hand, in quantum mechanics there are infinitely many elementary
measurements---corresponding to different observables---that provide
only partial information, whereas ``complementary'' informations could
be achieved only with mutually exclusive experiments where
non-commuting observables should be perfectly measured.

The problem then arises on how to achieve a kind of quantum
measurement that is ``complete'' \cite{prug,bus}, in the sense that it
can be used to infer information on all possible (also exclusive)
observables. The main idea is to perform a generalized ``unsharp''
measurement, described by a so-called POVM (positive-operator valued
measure), from which a specific type of information---i.e. a
particular ensemble average of a given operator---is retrieved by just
changing the data-processing of its experimental outcomes.

\par Informationally complete measurements are relevant for
foundations of quantum mechanics as a kind of ``standard'' for a
purely probabilistic description \cite{fuco}.  Moreover, the existence
of such measurements with minimal number of outcomes is crucial for
the quantum version of the de Finetti theorem \cite{caves}.

\par The most popular example of informationally complete measurement
is given by the coherent-state POVM for a single-mode of the radiation
field, whose probability distribution is the so-called $Q$-function
(or Husimi function) \cite{cah,notebalt,baltin}. Another example,
though of completely different kind, is the case of quantum tomography
\cite{tomo}, in which one measures an observable randomly selected
from an informationally complete set---a "quorum".

\par Investigations on informationally complete measurements have been
extensively carried out in the framework of "phase-space observables":
the monographs
\cite{Helstrom,Holevo,Davies,Busch,Schroeck,Hakioglu,Perinova} review
different aspects of these developments. Phase-space observables are
very useful in various fields of quantum physics, including quantum
communication and information theory. They also lead to substantial
advancement on some relevant conceptual issues, such as the problem of
jointly measuring non-commuting observables, or the problem of the
classical limit for quantum measurements.  However, the problem of
classifying all possible informationally complete measurements, also in
view of feasibility, was never investigated in generality, and only
elementary physical systems have been considered: the harmonic
oscillator and the spin.

\par In this paper, we present a more general treatment of the problem
based on group-theoretic techniques.  We will see that informationally
complete measurements can be achieved through POVM's derived from
unitary irreducible representations of a group on the Hilbert space of
the system. With the help of frame theory, we will also provide a
constructive way to evaluate the data-processing function for
estimating ensemble averages of arbitrary operators.  \par The paper
is organized as follows. In Sec. II we prove the equivalence of the
informational completeness of a measurement and the invertibility of
an operator constructed with the POVM. This proof makes use of frame
theory \cite{ds,czz}, and also shows how to obtain the data-processing
function for arbitrary operator. In Sec. III we derive the conditions
of informational completeness for POVM's that are covariant with
respect to a (compact) group that has unitary irreducible
representation on the Hilbert space of the system. We devote Sec. IV
to explicit examples of informationally complete POVM's with different
covariance group. In the example of the Weyl-Heisenberg group, we
recover the results of Ref. \cite{jmp}.  Some concluding remarks are
given in Sec. V.
\section{Info-complete POVM's and frame of operators}\label{gen}
In the following, we will make extensive use of the isomorphism
\cite{bellobs} between the Hilbert space of the Hilbert-Schmidt
operators $A,B$ on ${\cal H}$, with scalar product $\langle A,B\rangle
=\hbox{Tr}[A^\dag B]$, and the Hilbert space of bipartite vectors
$|A{\rangle\!\rangle},|B{\rangle\!\rangle}\in {\cal H}\otimes{\cal
H}$, with ${\langle\!\langle} A|B{\rangle\!\rangle}\equiv\langle
A,B\rangle $, and
\begin{eqnarray}
|A{\rangle\!\rangle}=\sum_{n=1}^d
\sum_{m=1}^d 
A_{nm}|n\rangle \otimes|m\rangle \;,\label{iso} 
\end{eqnarray}
where $|n\rangle$ and $|m\rangle$ are fixed orthonormal bases for
$\cal H$, $d=\hbox{dim }{\cal H}$, and $A_{nm}=\langle n|A|m \rangle
$. Notice the identities
\begin{eqnarray}
&&A\otimes B\dket{C}=\dket{ACB^\tau}\,,\nonumber \\
&&A\otimes B^\dag \dket{C}=\dket{ACB^*}\,,
\label{ids}
\end{eqnarray}
where $\tau$ and $*$ denote transposition and complex conjugation 
with respect to the fixed bases in Eqs. (\ref{iso}). 

\par An informationally complete measurement for a quantum system with
Hilbert space $\cal H$ is described by a POVM $\{\Pi_i\}$, $\Pi_i\ge 0$
and $\sum_i\Pi_i=I$, that allows to obtain the expectation
value of any operator $O$ of the system in the state $\rho $ as follows
\begin{eqnarray}
\langle O \rangle \equiv {\rm Tr} [\rho O]=
\sum_i f_i(O)\,{\rm Tr}[\rho\,\Pi_i]\,,
\label{def}
\end{eqnarray}
where $f_i(O)$ is the {\em data processing} function of the outcome
$i$ which depends on the operator $O$. Such a POVM will be referred
shortly to as "info-complete" POVM. Since Eq. (\ref{def}) holds for
any state $\rho $, it holds generally at the operator level without
the ensemble average, namely one has the expansion for operators 
\begin{eqnarray}
O=\sum_i f_i(O)\,\Pi_i\,.\;\label{exp}
\end{eqnarray}
Eq. (\ref{exp}) states that the set of positive operators $\Pi _i$
span the linear space of operators of the system. Spanning sets of
operators have been already used in quantum tomography \cite{macca}:
their characterization of spanning sets of operators is naturally
accomplished in the context of {\em frame theory} \cite{ds,czz}.  

\par An operator {\em frame} $\{\Xi_i\}$ is simply a set of operators
$\Xi_i$ that span a normed linear space of operators, namely there are
two constants $a$, and $b$ with $0<a\le b<\infty$ such that for all
operators $A$ one has $a|\!|A|\!|^2\le\sum_i |c_i(A)|^2\le
b|\!|A|\!|^2$, where $c_i(A)$ are the coefficients of the expansion of
$A$ over the set. Here, for simplicity, we will
consider the (Hilbert) space of Hilbert-Schmidt operators on ${\cal
  H}$, whence the norm will be the Frobenius
norm $|\!|A|\!|_2^2=\sqrt{\hbox{Tr}[A^\dag A]}$.
\par For $\{\Xi_i \}$ an operator frame there exists
another frame $\{\Theta _i \}$---called {\em dual frame}---giving the operator expansion in the form
\begin{eqnarray}
A=\sum _i {\rm Tr}[\Theta^\dag  _i A]\Xi _i\;.\label{sset}
\end{eqnarray}
The completeness relation of the frame and its dual reads
\begin{eqnarray}
\sum _i \langle \psi |\Xi _i |\phi \rangle \langle \varphi | \Theta
^\dag _i |\eta \rangle =\langle \psi |\eta \rangle \langle
\varphi|\phi \rangle \;, \label{ort}
\end{eqnarray}
for any $\phi, \varphi ,\psi ,\eta \in {\cal H}$.  For continuous
sets, the sums in Eqs. (\ref{sset}) and (\ref{ort}) are replaced by
integrals. 
Given a frame $\{\Xi_i \}$, generally the dual
set is not unique. However, all duals $\{\Theta_i\}$ of a given frame
can be obtained via the linear relation \cite{li}
\begin{eqnarray}
|\Theta_i\rangle\!\rangle=F^{-1}|\Xi_i\rangle\!\rangle+|Y_i\rangle\!\rangle-\sum_j
\langle\!\langle\Xi_j|F^{-1}|\Xi_i\rangle\!\rangle|Y_j\rangle\!\rangle\,,
\;\label{duals}
\end{eqnarray}
where $Y_i$ are arbitrary, and the positive and invertible operator
$F$ writes
\begin{eqnarray}
F=\sum_i|\Xi_i\rangle\!\rangle\langle\!\langle\Xi_i|\,.
\;\label{frame}
\end{eqnarray}
The operator $F$ is called "frame operator" in frame theory, 
whereas the set of operators corresponding to the vectors
$F^{-1}|\Xi_i\rangle\!\rangle$ through the above isomorphism is the so-called
"canonical dual" frame. On the other hand, given an arbitrary set of
operators $\{\Xi _i\}$, the invertibility of $F$ in Eq. (\ref{frame})
implies that the set is a frame.  Notice that if the frame is
bi-orthogonal, namely
$\langle\!\langle\Xi_i|F^{-1}|\Xi_j\rangle\!\rangle=\delta_{ij}$, then
the canonical one is the unique dual frame. One can also prove the
converse statement \cite{czz}, whence bi-orthogonality is
equivalent to uniqueness of the canonical dual frame.
\par From the above considerations it follows that a POVM $\{\Pi_i\}$
is info-complete if and only if the corresponding operator 
$F=\sum_i|\Pi_i\rangle\!\rangle\langle\!\langle\Pi_i|\,$ is
invertible. From linearity, by comparing Eq. (\ref{exp}) and
(\ref{sset}), one can see that a dual frame of an info-complete POVM
provides a data processing function as $f_i(O)=\hbox{Tr}[\Theta
^\dag _i O]$, whence Eq. (\ref{duals}) allows a useful flexibility in
the data-processing, with the possibility of minimizing the
statistical error of the estimation by varying the free operators $Y_i$.

\par Since the number of elements of an operator frame for $\cal H$
cannot be smaller than $d^2$, an info-complete POVM is necessarily not
orthogonal, whence it is overcomplete.  Viceversa, it is simple to
prove that an arbitrary frame for operators in $\cal H$ made of
positive operators $\{K_i\}$ allows to construct an info-complete
POVM. In fact, since the operator $S\equiv \sum _i K_i $ is
invertible, the set $\{\tilde K_i =S^{-1/2}K_i S^{-1/2} \}$ satisfies
the completeness relation $\sum _i \tilde K_i = I$.

\section{Group-theoretic techniques}
The representation theory of groups provides the easiest way to
construct frames made of unitary operators. Consider for example a unitary irreducible representation
(UIR) $\{U_g\,,\ g\in G\}$ of a compact group $G$ on the Hilbert space
$\cal H$. From the Shur's lemma, one has
\begin{eqnarray}
\int _{G} d\mu (g)\, U_g \,O\, U^\dag _g =
\hbox{Tr}[O] I\;,
\label{tro}
\end{eqnarray}
where $d\mu (g)$ denotes the left-invariant measure normalized as $
\int _G d\mu (g)=d$. As one can see, Eq. (\ref{tro}) is equivalent to
Eq. (\ref{ort}) with $\{U_g\}$ self-dual operator frame. On
the other hand, the direct construction of info-complete POVM's is not
as simple, since it involves the searching of frames of {\em
positive} operators.  A way to construct info-complete POVM's is
suggested by Eq. (\ref{tro}).  For any density matrix $\nu $ the set
of positive operators
\begin{eqnarray}
\Pi _g= U_g \nu U^\dag _g\;\label{10}
\end{eqnarray}
provides a resolution of the identity,  
whence  $\{\Pi _g \}$ is a POVM. Moreover, the POVM is info-complete 
iff the operator 
\begin{equation}
F=\int _G d\mu (g) \, \dket{\Pi _g} \dbra{\Pi_g }
=\int _G d\mu (g)\,U_g \otimes U_g ^* \dket{\nu } \dbra{\nu }
U_g^\dag  \otimes (U_g ^*)^\dag 
\;,\label{inte}
\end{equation}
is invertible, where we used Eq. (\ref{ids}). Representation theory
allows to evaluate the integral in Eq. (\ref{inte}). When $U_g \otimes
U_g^*$ has only inequivalent irreducible representations on $\cal H
\otimes \cal H$, upon denoting by $P_\sigma $ the projectors over the
invariant subspaces, one has
\begin{eqnarray}
F= d\sum _\sigma \frac{\hbox {Tr}[P_\sigma \dket{\nu }\dbra {\nu
}]}{\hbox{Tr}[P_\sigma ]}\,P_ \sigma \;.\label{framop}
\end{eqnarray}
As a consequence, the POVM $\{\Pi _g\}$ is info-complete iff the state
$\nu$ satisfies the condition 
\begin{eqnarray}
\hbox {Tr}[P_\sigma \dket{\nu }\dbra {\nu}]\neq0\,\quad\forall\sigma\,.
\label{invcond}
\end{eqnarray}
In this case the inverse of $F$ is easily calculated as follows 
\begin{eqnarray}
F^{-1}=d^{-1}\sum _\sigma \frac{\hbox{Tr}[P_\sigma ]}{\hbox
{Tr}[P_\sigma \dket{\nu }\dbra {\nu }]}\,P_ \sigma \;,\label{framinv}
\end{eqnarray}
and the canonical dual $\Theta _g$  is obtained by the identity 
$\dket{\Theta _g}=F^{-1}\dket{U_g\nu U_g^\dag}$, namely 
\begin{eqnarray}
\dket{\Theta_g}&=&d^{-1}\sum _\sigma \frac{\hbox{Tr}[P_\sigma ]}{\hbox
{Tr}[P_\sigma \dket{\nu }\dbra {\nu }]}\,P_\sigma\dket{U_g\nu
U_g^\dag}\nonumber\\ &=& d^{-1}U_g\otimes U^*_g \sum _\sigma
\frac{\hbox{Tr}[P_\sigma ]}{\hbox {Tr}[P_\sigma \dket{\nu }\dbra {\nu
}]}\,P_\sigma\dket{\nu}\;,\label{candu}
\end{eqnarray}
where we used the property $[U_g\otimes U^*_g, P_\sigma ]=0$.  By
Eq. (\ref{candu}) one can notice that the canonical dual is covariant
itself, namely $\Theta _g= U_g \xi U_g^\dag $, where $\xi $ is
given by 
\begin{eqnarray}
\dket{\xi}=d^{-1}\sum _\sigma
\frac{\hbox{Tr}[P_\sigma ]}{\hbox {Tr}[P_\sigma \dket{\nu }\dbra {\nu
}]}\,P_\sigma\dket{\nu}\;.
\end{eqnarray}
At this stage we can make some general remarks. Among the invariant
subspaces there is always the span of $\frac1{\sqrt d}\dket{I}$, and
thus Eq. (\ref{candu}) has always the term
\begin{eqnarray}
d^{-1} U_g\otimes U^*_g \,\dket{I}=d^{-1}\,\dket{I}\,.
\end{eqnarray}
The other invariant subspaces depend on the representation $U_g$, but
we can prove that for any UIR $U_g$ such that $U_g\otimes U^*_g$ has
inequivalent irreducible representations, there always exists a
suitable $\nu$ such that the POVM $U_g\nu U^\dag_g$ is
info-complete. In fact, upon writing the projectors $P_\sigma $ in
terms of their eigenvectors 
\begin{eqnarray}
P_\sigma=\sum_j\dket{\Psi^{(\sigma)}_j}\dbra{\Psi^{(\sigma)}_j}\,,
\end{eqnarray}
from $\sum_\sigma P_\sigma=I$, it follows that $\{\Psi_j^{(\sigma)}\}$
is an orthonormal basis for the Hilbert-Schmidt operators. By
identifying $\Psi^{(0)}_0 \equiv \frac I{\sqrt d}$, from the
orthogonality one has $\hbox{Tr}[\Psi ^{(\sigma )}_j]=\sqrt d \,\delta
_{\sigma 0}$. Then, one can find suitable phases $\{\theta _\mu \}$
and a real constant $\alpha $ such that the Hermitian operator
\begin{eqnarray}
\nu =\frac I d + \alpha \sum _{\mu \neq 0}(e^{i\theta _\mu }\Psi
^{(\mu )}_ j+e^{-i\theta _\mu }{\Psi
^{(\mu )}_ j}^\dag )
\;\label{nugud}
\end{eqnarray}
is a density matrix satisfying  condition (\ref{invcond}). Notice that
in Eq. (\ref{nugud}) we just need a single label $j=j(\mu)$ for each $\mu $.   

\par In the last part of this Section, we want to notice that all
POVM's of the form (\ref{10}) are equivalent to a generalized Bell
measurement \cite{bellobs} on a tensor-product space ${\cal H}\otimes
{\cal H}$, where the second space describes an ancilla prepared in the
state $\nu ^\tau$. In fact, one has
\begin{eqnarray}
U_g \nu U^\dag _g =\hbox{Tr}_{\cal A}[(I\otimes \nu^\tau)\dket{U_g}\dbra{U_g}]\;, 
\label{trp}
\end{eqnarray}
where $\hbox{Tr}_{\cal A}$ denotes the partial trace over the ancilla
space.  In general, the projectors on the maximally entangled states
$\dket{U_g}$ are not orthogonal. The above considerations allow to
understand the construction of the quantum universal detectors
introduced in Ref. \cite{univdet}.
 
\section{Examples} 
In this section we will provide some examples of info-complete POVM's,
showing their underlying group structure.

\subsection{${\mathbb Z}_d\times{\mathbb Z}_d$}
This first example involves a minimal info-complete POVM, namely a
POVM having $d^2$ elements, and will give us some general insight in
the case of projective representations of abelian groups.  Consider
the group ${\mathbb Z}_d\times{\mathbb Z}_d$, and the following
$d$-dimensional projective UIR
\begin{eqnarray}
U_{m,n}=\sum _{k=0}^{d-1} e^{\frac {2\pi i}{d}km}|k \rangle \langle k
\oplus n|\;,
\label{ZZ}
\end{eqnarray}
where $m,n\in [0,d-1]$, and $\oplus $ denotes sum modulo $d$.  The
composition and orthogonality relations of the set are given by
\begin{eqnarray}
&&U_{m,n}\,U_{p,q}\,U^\dag
  _{m,n}=e^{\frac{2\pi i}{d}(np-mq)}\,
  U_{p,q}\;,  \label{propUs1} \\& & 
\hbox{Tr}\left[U^\dag_{p,q}\; U_{m,n}\right] = d\,\delta_{mp}\: \delta_{nq} 
\label{propUs2}\;.
\end{eqnarray}
We will now look for an info-complete POVM of the form
\begin{eqnarray}
\Xi_{m,n}=\frac1dU_{m,n}\,\nu \,U^\dag_{m,n}\,.
\label{zdzd}
\end{eqnarray}
The properties of the projective UIR help us to find the conditions for
informational completeness, and to evaluate the dual frame directly, as an
alternative way to the general method developed in the previous section.
First, let us expand the state $\nu$ in Eq. (\ref{zdzd}) on the basis
of $\{U_{m,n} \}$,
\begin{eqnarray}
\Xi_{m,n}=\frac1d\sum_{p,q}e^{\frac{2\pi
i}d(np-mq)}\,\hbox{Tr}[U^\dag_{p,q}\nu]\,U_{p,q}\;,
\end{eqnarray}
where we used Eq. (\ref{propUs1}). From the identity
    $\sum_{n}e^{\frac{2\pi i}dnp}=d\,\delta_{p0}$,
    one easily checks that a dual frame for the POVM $\Xi _{m,n}$ is
    given by
\begin{eqnarray}
\Theta_{m,n}=\frac1d\sum_{p,q}e^{\frac{2\pi
i}d(np-mq)}\,\frac{U_{p,q}}{\hbox{Tr}[U_{p,q}\nu]}
\;,
\label{dualzdzd}
\end{eqnarray}
Then, the condition for informational completeness on $\nu$ is simply 
\begin{eqnarray}
\hbox{Tr}[U^\dag_{p,q}\nu]\neq0\,,\quad\forall (p,q)\in{\mathbb
Z}_d\times{\mathbb Z}_d\,.
\end{eqnarray}
A pure state $\nu =|\psi \rangle \langle \psi |$ that satisfies the
above condition is given by 
\begin{eqnarray}
|\psi\rangle=
\sqrt{\frac{1-|\alpha|^2}{1-|\alpha|^{2d}}}\sum_{n=0}^{d-1}\alpha^n|n\rangle\;,
\end{eqnarray}
for any $\alpha$ with $0<|\alpha|<1$.  Notice that the sets $\Theta
_{m,n}$ and $\Xi _{m,n}$ are biorthogonal, hence $\Theta _{m,n}$ is
the unique dual set.  The results given in this example are consistent
with the general treatment of Sec. \ref{gen}. In fact, the irreducible
representations of $U_{m,n}\otimes U_{m,n}^*$ are all inequivalent and
one-dimensional, and the invariant subspaces are just the spans of
$U_{m,n}$'s.

\subsection{$SU(d)$}
In this second example we will examine the POVM
\begin{eqnarray}
\frac1d\,U_g\nu U^\dag_g\,,\qquad U_g\in SU(d)\,.
\end{eqnarray}
Here the invariant subspaces of $U_g\otimes U^*_g$ are the
linear span of $\frac1{\sqrt{d}}\dket{I}$ and its orthogonal
complement. From Eq. (\ref{framop}), the frame operator can be
expressed as
\begin{eqnarray}
F=\frac1d\dket{I}\dbra{I}+\frac{d\hbox{Tr}
[\nu^2]-1}{d^2-1}\left(I-\frac1d\dket{I}\dbra{I}\right)\,,
\end{eqnarray}
and it is invertible iff
\begin{eqnarray}
d\hbox{Tr}[\nu^2]-1\neq0\,.
\end{eqnarray}
Notice that $\hbox{Tr}[\nu^2]=\frac1d$ only for $\nu=\frac Id$, which
leads to a trivial POVM. Any other $\nu$ gives an info-complete
POVM. The inverse of the frame operator writes 
\begin{eqnarray}
F^{-1}=\frac1d\dket{I}\dbra{I}+
\frac{d^2-1}{d\hbox{Tr}[\nu^2]-1}\left(I-\frac1d\dket{I}\dbra{I}\right)\,.
\end{eqnarray}
Finally, by Eq. (\ref{candu}), the canonical dual reads
\begin{eqnarray}
\Theta _g =
U_g\left(\frac{d^2-1}{d\hbox{Tr}[\nu^2]-1}\nu-\frac{d-\hbox{Tr}[\nu^2]}{d\hbox{Tr}[\nu^2]-1}I\right)U^\dag_g\,.
\end{eqnarray}
In Ref. \cite{univproc} we showed that this canonical dual is optimal
among all covariant duals, for estimating expectations of arbitrary
Hermitian operators with minimal r.m.s error.

\subsection{Weyl-Heisenberg group} 
The last example involves an infinite dimensional system, and is
somehow the continuous counterpart of the first example. We will
consider the Weyl-Heisenberg group in the representation of
displacements $D(\alpha)=e^{\alpha a^\dag-\alpha^* a}$, where $a$ and
$a^\dag$ are the annihilation and creation operators of a boson field,
i.e. $[a,a^\dag ]=1$. Notice that the group is not compact, hence the
general treatment of Sec. \ref{gen} does not directly apply. However,
the displacement representation is square integrable \cite{gmp}, and
the main identity (\ref{tro}) still holds in the form
\begin{eqnarray}
\int _{\mathbb C}\frac{d^2 \alpha  }{\pi }D(\alpha ) \,O\,D^\dag
(\alpha )= \hbox{Tr}[O] I \;.\label{stil}
\end{eqnarray}
The group structure is revealed by the following identities
\begin{eqnarray}
D(\alpha)D(\beta)D^\dag
(\alpha)=e^{\alpha\beta^*-\alpha^*\beta}\;D(\beta)\label{phaseds}\,,\\
\hbox{Tr}[D^\dag (\alpha)
D(\beta)]=\pi\,\delta ^{(2)}(\alpha-\beta)\;, \label{orthogds}
\end{eqnarray}
where $\delta ^{(2)}(\alpha )\equiv (1/ \pi ^2)\int_{\mathbb C}d^2
\gamma \, e^{\alpha \gamma ^*-\alpha ^* \gamma }$ denotes the
Dirac-delta on the complex plane.  From Eqs. (\ref{stil}) and
(\ref{orthogds}) , it also follows the completeness on ${\cal
H}\otimes {\cal H}$
\begin{eqnarray}
\int _{\mathbb C}\frac {d^2 \alpha }{\pi }
\dket{D(\alpha )} \dbra{D(\alpha )} = I \otimes I \;,
\end{eqnarray}
and the orthogonality in Dirac sense
\begin{eqnarray}
\dbra{D(\alpha )} D(\beta )\rangle\!\rangle =\pi \delta ^{(2)}(\alpha
-\beta )
\;.
\end{eqnarray}
We consider the POVM
\begin{eqnarray}
\Pi (\alpha)=\frac 1\pi D(\alpha)\nu D^\dag (\alpha)\,,\label{povmds}
\end{eqnarray}
where $\nu $ is an arbitrary normalized state. By expanding 
$\nu $ as $\nu =\int_{\mathbb C}\frac {d^2 \gamma }{\pi }
\hbox{Tr}[D^\dag (\gamma)\nu]D(\gamma )$ and using Eq. (\ref{phaseds}), one has 
\begin{eqnarray}
\Pi (\alpha)=\int_{\mathbb C}\frac {d^2 \gamma }{\pi }\,
e^{\alpha\gamma^*-\alpha^*\gamma}\,
\hbox{Tr}[D^\dag (\gamma)\nu]D(\gamma)\;,
\end{eqnarray}
The frame operator can then be written as follows
\begin{eqnarray}
F&=&\int_{\mathbb C}\frac{d^2 \alpha }{\pi }\int_{\mathbb C}\frac{d^2
\beta }{\pi }\int_{\mathbb C}\frac{d^2 \gamma }{\pi } 
\,e^{\alpha(\beta^*-\gamma^*)-\alpha^*(\beta-\gamma)}
\nonumber \\&\times  & 
\hbox{Tr}[D^\dag (\beta)\nu] \hbox{Tr}[D^\dag (\gamma )\nu ]^*
\dket{D(\beta)}\dbra{D(\gamma)}\nonumber\\
&=&\int_{\mathbb C}\frac{d^2 \beta }{\pi }|\hbox{Tr}[D^\dag
  (\beta)\nu]|^2\,
\dket{D(\beta)}\dbra{D(\beta)}\,.
\end{eqnarray}
The POVM is then info-complete iff $\hbox{Tr}[D(\beta)^\dag\nu]\neq0$
for all $\beta$. We notice that such a condition was also found in
Ref. \cite{jmp} in the context of phase-space representation and
covariant localization observables.  

\par\noindent The inverse of the frame operator writes
\begin{eqnarray}
F^{-1}=\int_{\mathbb C}\frac{d^2 \beta }\pi 
\frac1{|\hbox{Tr}[D^\dag (\beta)\nu]|^2}\dket{D(\beta)}\dbra{D(\beta)}\,.
\end{eqnarray}
The canonical dual can be finally evaluated using Eq. (\ref{candu}),
and is given by
\begin{eqnarray}
\Theta(\alpha)=D(\alpha )\left (\int_{\mathbb C}\frac{d^2 \beta }\pi \,
\frac {D(\beta) }{\hbox{Tr}[D (\beta)
\nu]}\right )D^\dag (\alpha )\;.\label{het}
\end{eqnarray}
Notice that the dual is unique since it can be readily checked that
the POVM and the canonical dual are biorthogonal.  \par The present
example in infinite dimension needs some care in checking the convergence
of the processing function $f(\alpha ,O)=\hbox{Tr}[\Theta ^\dag
  (\alpha )O] $. In fact, if we take the vacuum
state $\nu=|0\rangle\langle 0|$, the POVM in Eq. (\ref{povmds}) will
be reduced to the customary projection on coherent states $|\alpha
\rangle $, and the
measurement will correspond to phase-space averaging with the 
$Q$-function $Q(\alpha )=\frac{1}{\pi}\langle \alpha
|\rho|\alpha \rangle$. We know that this gives expectations
only for operators admitting anti-normal ordered field expansion
\cite{baltin}.  In particular, the matrix elements of the density
operator cannot be recovered in this way \cite{tomo}. Therefore, in
infinite dimensions the universality can be limited by convergence.

\section{Conclusions}
We have presented a group-theoretical method to construct informationally
complete quantum measurements. This method
allows to find the conditions for such completeness, and to construct
a wide class of info-complete POVM's from unitary irreducible
representation of groups. These POVM's can always be viewed as
projectors on maximally entangled states---generally not
orthogonal---of the system coupled with an ancilla, thus relating
info-complete POVM's with the quantum universal detectors of
Ref. \cite{univdet}. The processing functions pertaining any arbitrary
operator have been obtained using the general method of frame
theory. Such functions are generally not unique, and this allows
optimizing the frame in order to minimize the statistical error of the
estimation.  We have finally provided some examples of info-complete POVM's
corresponding to different kind of groups: discrete and abelian
(${\mathbb Z}_d\times{\mathbb Z}_d$), continuous and compact
($SU(d)$), and non compact and abelian (Weyl-Heisenberg).  

\subsection*{Acknowledgments}
This work has been cosponsored by EEC through the ATESIT project
IST-2000-29681 and by MIUR through Cofinanziamento-2002. P. P. and
M. F. S. also acknowledge support from INFM through the project
PRA-2002-CLON.


\begin{thebibliography}{99}
\bibitem{notebalt} However, because of convergence problems this
function gives only expectation values of operators that admit
anti-normal ordered expansion \cite{baltin}.
\end{thebibliography}
\end{document}